\documentclass[trackchanges]{aastex701}

\usepackage{booktabs}
\usepackage{graphicx}
\usepackage{subcaption}
\usepackage{xcolor}
\usepackage{amsmath}
\usepackage{CJKutf8}
\usepackage{soulutf8}
\setstcolor{red}

\usepackage[dvipsnames]{xcolor}

\newcommand{\localname}[1]{%
  \begin{CJK*}{UTF8}{mj}#1\end{CJK*}%
}

\begin{document}




\title{Toward Operational Solar Flare Peak Flux Nowcasting: A Strategy Combining Real-Time Data, Machine Learning, and NOAA Flare Detection Criteria}


\correspondingauthor{Bo Shen}
\email{bo.shen@njit.edu}

\author[orcid=0000-0003-4342-9483]{Kangwoo Yi (\localname{이강우})}
\affiliation{Department of Physics, New Jersey Institute of Technology, University Heights, Newark, NJ 07102, USA}
\email{ky263@njit.edu}  

\author[orcid=0000-0000-0000-0002]{Qin Li}
\affiliation{Department of Physics, New Jersey Institute of Technology, University Heights, Newark, NJ 07102, USA}
\email{ql47@njit.edu}

\author[orcid=0000-0001-6460-408X]{Haodi Jiang}
\affiliation{Department of Computer Science, Sam Houston State University, Huntsville, TX 77341, USA}
\email{hxj024@shsu.edu}

\author[orcid=0000-0002-2633-3562]{Meiqi Wang}
\affiliation{Department of Physics, New Jersey Institute of Technology, University Heights, Newark, NJ 07102, USA}
\email{mw335@njit.edu}

\author[orcid=0000-0000-0000-0002]{Haimin Wang}
\affiliation{Department of Physics, New Jersey Institute of Technology, University Heights, Newark, NJ 07102, USA}
\affiliation{Big Bear Solar Observatory, New Jersey Institute of Technology, Big Bear City, CA 92314, USA}
\email{haimin.wang@njit.edu}

\author[orcid=0000-0002-2643-3600]{Bo Shen} 
\affiliation{Department of Mechanical and Industrial Engineering, New Jersey Institute of Technology, University Heights, Newark, NJ 07102, USA}
\affiliation{Department of Data Science, New Jersey Institute of Technology, University Heights, Newark, NJ 07102, USA}
\email{bo.shen@njit.edu}

\begin{abstract}
We present the RMN strategy (Real-time data, machine learning, and NOAA flare detection criteria) for nowcasting the peak soft X-ray flux of ongoing solar flares under operationally realistic conditions. 
The strategy combines real-time GOES 0.1–0.8 nm X-ray observations with an attention-based sequence-to-sequence Long Short-Term Memory model. Under the NOAA flare detection criteria, predictions are evaluated at one-minute intervals from three minutes after the cataloged onset to the observed peak using the preceding 60 minutes of X-ray observations. 
We apply the RMN strategy to C-, M-, and X-class flares observed by GOES-8–18 from 1997 to 2024 using four-fold cross-validation. 
The major results of this study are as follows.
First, the model nowcasts peak soft X-ray flux with RMSE and PE values of 0.26 and 3.11\% for the $\geq$C-class group, 0.45 and 5.59\% for the $\geq$M-class group, and 0.87 and 12.76\% for the X-class group. 
The higher discrepancy toward stronger flare groups indicates that peak-flux prediction is more challenging for higher-intensity flares.
Second, the model performance depends on flare rise time and prediction time, with larger errors for longer rise time events and improved performance as the prediction time approaches the flare peak. 
Shorter rise time events approach their final peak more rapidly, providing a clearer indication of the eventual peak, whereas the larger difference for longer rise time events may partly reflect more complex temporal evolution.
Third, empirical coverage based on total uncertainty remains high but decreases for stronger flares, with noise uncertainty contributing more than model uncertainty.


\end{abstract}



\section{Introduction} \label{sec:intro}
Solar flares are sudden releases of magnetic energy in the solar atmosphere, producing intense electromagnetic radiation over a wide spectral range from radio waves to X-rays and gamma rays \citep{1984Natur.312..623R, 2002A&ARv..10..313P}. In addition to radiation, large flares can be associated with energetic particles and coronal mass ejections, thereby contributing to broader space weather disturbances \citep{1979SoPh...61..201M, 1992ARA&A..30..113K, 2013SSRv..175...53R, 2018ApJ...869...99K}. 
Because flare radiation reaches Earth at nearly the speed of light, its impacts on the geomagnetosphere can occur almost immediately after the flare begins. 
In particular, enhanced soft X-ray emission increases ionization in the ionosphere and can affect high-frequency radio communication, GPS navigation, satellite operations, and aviation \citep{2000JASTP..62.1223S, 2014Life....4..491C, 2018ApJ...856....7H, 2020JSWSC..10...27C}.
These operational impacts make the timely estimation of flare intensity an important component of space weather prediction.

The Geostationary Operational Environmental Satellite (GOES) observes solar X-ray fluxes in the 0.1–0.8 nm channel, which is widely used for flare detection, classification, and alerting. 
During a flare, the soft X-ray flux typically increases rapidly during the rise phase, reaches a maximum at the flare peak time, and then decays during the gradual phase. 
Because space weather hazards are associated with flare intensity, estimating the eventual peak X-ray flux as early as possible is important, although the peak flux remains unknown during the rise phase and must be inferred from the observations available during the early evolution of the flare. 
Most solar flare prediction studies have focused on forecasting flare occurrence and the maximum flare class within a future time window, typically within the next 24 hr \citep{1990SoPh..125..251M, 2012ApJ...747L..41B, 2015ApJ...798..135B, 2018ApJ...869...91P, 2019JKAS...52..133L, 2021ApJ...910....8Y}. In contrast, flare nowcasting aims to predict solar flare activity on a timescale of minutes, providing rapid estimates as the flare evolves.


Several previous studies have explored ideas related to flare nowcasting.
\citet{2019SpWea..17.1783G} developed a nowcasting approach for X-class flares using perturbations in very-low-frequency (VLF) radio-wave propagation, showing that VLF phase measurements can be used to estimate long-wavelength X-ray fluxes during large flares. 
\citet{2020ApJ...890L...5Y} used GOES X-ray data and sequence-to-sequence \citep[seq2seq;][]{seq2seq_2014} deep-learning models to forecast $\geq$ M-class flare flux evolution during the rise phase, showing that deep-learning methods can predict soft X-ray profiles more effectively than several conventional regression models. 
That work demonstrated the feasibility of learning flare-profile evolution directly from X-ray time series data. 
\citet{2025ApJ...993...95T} investigated a flare nowcasting method using Hot Onset Precursor Event (HOPE) signatures derived from GOES X-ray data. Their HOPE-based method predicts $\geq$ C5.0 class flare trigger before flare peak time. 
This provides a demonstration that early flare-phase plasma analysis can improve flare nowcasting performance. 

In this study, we present the RMN strategy, which combines Real-time data, Machine learning models, and the National Oceanic and Atmospheric Administration (NOAA) flare detection criteria, to nowcast the peak soft X-ray flux of ongoing solar flares.
Specifically, we obtain GOES X-ray observation data and a seq2seq framework using Long Short-Term Memory (LSTM; \citealp{Hochreiter1997LSTM}) and an attention mechanism \citep{Bahdanau2014NeuralMT}.
We apply the RMN strategy to $\geq$ C1.0 flares observed from 1997 to 2024 and evaluate solar flare nowcasting performance for predicting the peak X-ray flux of ongoing solar flares under realistic operational conditions.

This paper is organized as follows. The data are described in Section~\ref{sec:data}. The deep-learning model and evaluation methods are described in Section~\ref{sec:methods}. Our flare nowcasting performance is presented in Section~\ref{sec:results}. Conclusion and discussion are presented in Section~\ref{sec:conclusion_discussion}.

\section{Data} \label{sec:data}
In this study, we use C-, M-, and X-class solar flare events observed by GOES-8 through GOES-18 from 1997 to 2024\footnote{\url{ftp://ftp.swpc.noaa.gov/pub/indices/}}. 
The event list includes the start, peak, and end times of each flare, the observed GOES, and the flare class determined from the peak soft X-ray flux. 
Flares with neighboring peak times separated by less than one hour are treated as multiple flare events and excluded from the dataset.

For each event, we use the one-minute X-ray flux in the 0.1–0.8 nm channel measured by the GOES instrument listed as the observing instrument in the flare catalog. 
To combine X-ray data from the GOES-8-15 and GOES-R series (GOES-16-18), we use the Science-Quality GOES-8–15 X-ray data together with the GOES-R series data. 
The original operational GOES-8–15 X-ray data include SWPC scaling factors to match GOES-7 and are affected by discontinuities from sporadic calibration updates and differences in bandpass assumptions. 
In contrast, GOES-16-18 X-ray data are provided in true physical units without SWPC scaling. 
The Science-Quality GOES-8–15 X-ray data are reprocessed to remove these scaling factors and incorporate calibration and bandpass corrections, making them more suitable for long-term analyses alongside GOES-R observations\footnote{\url{https://www.ncei.noaa.gov/data/goes-space-environment-monitor/access/science/xrs/GOES_1-15_XRS_Science-Quality_Data_Readme.pdf}}. 
Since solar X-ray flux spans several orders of magnitude from weak to strong flares, all X-ray flux values are converted to a logarithmic scale before being used as model input.


The highly imbalanced occurrence of X-, M-, and C-class flares (approximately 1{:}10{:}100) limits the number of X-class flares available for model evaluation in a fixed training--test split \citep{2023ApJS..265...34Y}. 
This limited test sample can make the estimated performance for high-intensity events strongly dependent on which rare events are assigned to the test set. 
To evaluate the model over all available strong flares, we use four-fold cross-validation. The X-ray data are divided into four folds according to the month of flare occurrence, with each fold covering a three-month interval. Specifically, we have Fold 1: January--March; Fold 2: April--June; Fold 3: July--September; and Fold 4: October--December.  The model is trained and tested four times, with one fold used as the test set and the remaining three folds used as the training set.
The final performance scores are obtained by averaging the scores from the four test sets.
Table~\ref{tab:cv_numberofflare} shows the number of flare events included in the dataset.

\begin{table}[!htb]
\centering
\caption{
Number of flare events in each fold of our four-fold cross-validation. 
}
\label{tab:cv_numberofflare}
\hspace{-0.3in}\begin{tabular}{lrrrr}
\toprule
\textbf{Dataset} & \textbf{C-class} & \textbf{M-class} & \textbf{X-class} & \textbf{Total} \\
\midrule
Fold 1 & 3810 (87.0\%) & 527 (12.0\%) & 41 (0.9\%) & 4378 \\
Fold 2 & 4026 (85.9\%) & 603 (12.9\%) & 57 (1.2\%) & 4686 \\
Fold 3 & 4178 (83.7\%) & 752 (15.1\%) & 61 (1.2\%) & 4991 \\
Fold 4 & 4554 (84.6\%) & 762 (14.2\%) & 67 (1.2\%) & 5383 \\
\midrule
Total  & 16568 (85.2\%) & 2644 (13.6\%) & 226 (1.2\%) & 19438 \\
\bottomrule 
\end{tabular}\hspace{+0.3in}
\end{table}

We construct the dataset using a sliding-window approach applied to the X-ray flux profile of each flare event. 
For GOES-8--15 observations, the NOAA flare detection algorithm defined the flare start time as the first of four consecutive one-minute XRS-B measurements satisfying the prescribed rise conditions, requiring three additional minutes of observations to identify the flare \citep{{2016A&A...592A.133R}}. 
For GOES-16 and later observations, the GOES-R XRS flare detection algorithm identifies flare start time using multiple conditions applied to a nine-minute real-time data frame and retrospectively determines the flare start time from the preceding measurements\footnote{\url{https://data.ngdc.noaa.gov/platforms/solar-space-observing-satellites/goes/goes16/l2/docs/GOES-R_XRS_L2_Data_Users_Guide.pdf}}. 
Thus, under both procedures, the flare start time is determined only after subsequent X-ray measurements become available. 
To standardize the model evaluation across the GOES-8--18 period, we assume that each flare is first detected three minutes after its cataloged start time and begin peak-flux prediction from this assumed recognition time. 
For each event, prediction times are assigned at one-minute intervals from the assumed detection time to the observed peak time. 
For each prediction time, the input data consists of the preceding 60 minutes of 0.1--0.8 nm X-ray flux, and the prediction target is the peak X-ray flux of the ongoing flare.
Table~\ref{tab:cv_numberofdata} presents the number of X-ray time series data constructed from these events.  These samples represent the actual data composition used for model development.

\begin{table}[!htb]
\centering
\caption{
Number of 60-minute sliding-window X-ray time series data in each fold of our four-fold cross-validation. 
}
\label{tab:cv_numberofdata}
\hspace{-0.3in}\begin{tabular}{lrrrr}
\hline
\textbf{Dataset} & \textbf{C-class} & \textbf{M-class} & \textbf{X-class} & \textbf{Total} \\
\hline
Fold 1 & 38105 (80.2\%) & 8513 (17.9\%) & 917 (1.9\%) & 47535 \\
Fold 2 & 41438 (80.0\%) & 9055 (17.5\%) & 1285 (2.5\%) & 51778 \\
Fold 3 & 42545 (75.5\%) & 12611 (22.4\%) & 1231 (2.2\%) & 56387 \\
Fold 4 & 42875 (76.8\%) & 11417 (20.4\%) & 1560 (2.8\%) & 55852 \\
\hline
Total  & 164963 (78.0\%) & 41596 (19.7\%) & 4993 (2.4\%) & 211552 \\
\hline
\end{tabular}\hspace{+0.3in}
\end{table}




\section{Methods} \label{sec:methods}

\subsection{Machine Learning Model} \label{subsec:modeling}

To evaluate the flare nowcasting capability of the RMN strategy, we construct a deep-learning model to estimate the peak soft X-ray flux of an ongoing solar flare from its early X-ray time series. 
The model follows a seq2seq framework with LSTM layers and an attention mechanism.
LSTM is a type of recurrent neural network designed to capture temporal dependencies in the X-ray flux sequence. The seq2seq framework consists of two LSTM networks. The encoder processes the input sequence and extracts its information, while the decoder uses this representation to generate the forecast.
The attention mechanism allows the decoder to selectively use informative encoder states for the final prediction.

\begin{figure}
\centering
\includegraphics[width=0.5\linewidth]{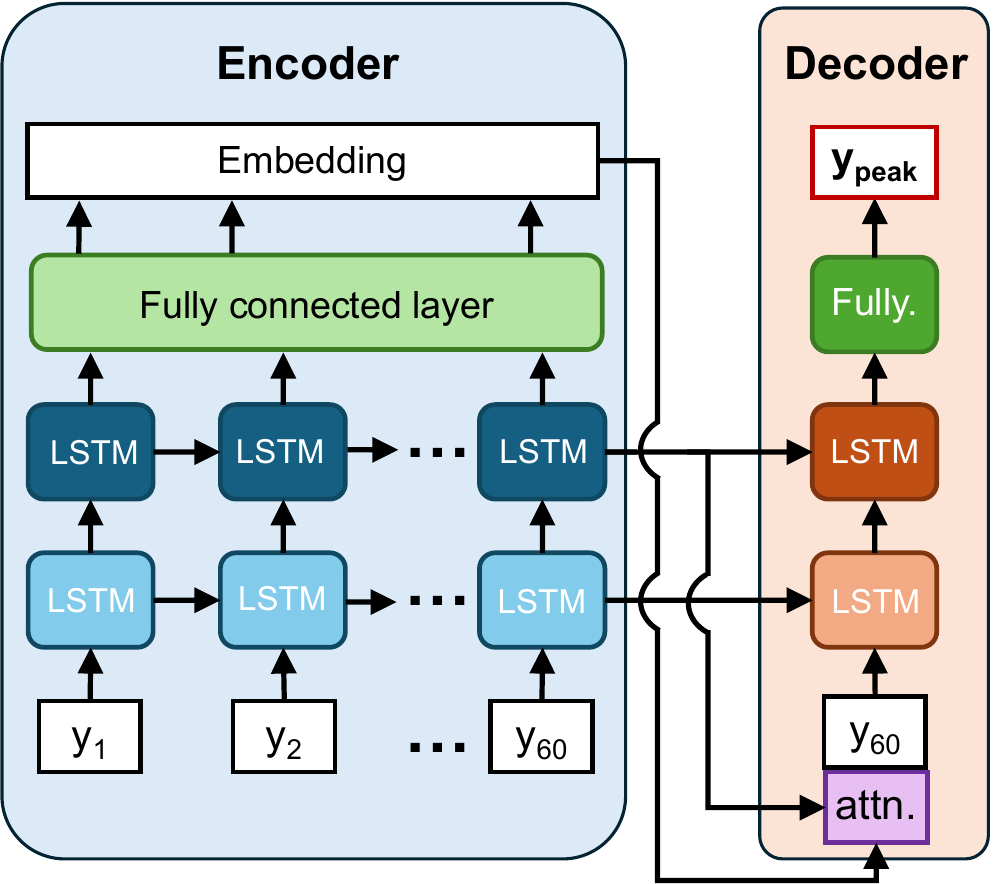}
\caption{
Seq2seq with attention model structure. The encoder and decoder consist of two LSTM layers and one fully connected layer with 1024 nodes. The embedding size is 8. The output $y_{peak}$ is the predicted peak X-ray flux. 
}
\label{fig:model_structure}
\end{figure}

Figure~\ref{fig:model_structure} illustrates the proposed model architecture. 
The encoder takes a 60-minute X-ray flux sequence, i.e, $y_{1}, y_{2},...,y_{60}$, as input. 
The input sequence is processed by two LSTM layers, and the resulting hidden states are passed through a fully connected layer to generate an embedded representation of the flare features. 
In the decoder, an attention layer takes the embedded features and the final encoder hidden state, and produces a context vector as a weighted summary of the embedded features.
The context vector is concatenated with the final observed flux value, $y_{60}$, before being passed through two decoder LSTM layers and a fully connected layer to produce a scalar value, $y_{peak}$, representing the predicted peak X-ray flux of the ongoing flare.
All LSTM and fully connected layers have 1024 nodes. 
The model is trained using the mean squared error loss function. 
We use the Adam optimizer \citep{Kingma:2014vow} with a learning rate of 0.001.

\subsection{Evaluation Method} \label{subsec:evaluation}

To evaluate the model performance, each performance metric is computed separately for the test set of each fold, and the final value is obtained by averaging the four fold-wise scores. 
We use Root Mean Square Error (RMSE; \citealp{Hodson2022}), Percentage Error (PE; \citealp{DEMYTTENAERE201638}), and the Pearson Correlation Coefficient (CC; \citealp{Rodgers1988}) to evaluate the peak flux prediction performance. Since the target peak flux spans several orders of magnitude, the regression metrics are computed using logarithmic flux values unless otherwise stated.

The RMSE is defined as
\begin{equation}
    \mathrm{RMSE}
    =
    \sqrt{
    \frac{1}{N}
    \sum_{i=1}^{N}
    \left(
    y_i - \hat{y}_i
    \right)^2
    } ,
    \label{eq:rmse}
\end{equation}
where $N$ is the number of test samples, $y_i$ is the observed peak soft X-ray flux of the $i$th sample, and $\hat{y}_i$ is the corresponding predicted peak flux. 
This metric measures the average magnitude of the prediction error in logarithmic flux space. 
RMSE measures the average magnitude of the prediction error, with a smaller value indicating better model performance.

The PE is defined as
\begin{equation}
    \mathrm{PE}
    =
    \frac{1}{N}
    \sum_{i=1}^{N}
    \left|
    \frac{y_i - \hat{y}_i}{y_i}
    \right|
    \times 100 \% ,
    \label{eq:percentage_error}
\end{equation}
which calculates the average relative difference between the observed and predicted peak flux, with lower values indicating better results.

The CC is defined as
\begin{equation}
    CC
    =
    \frac{
    \sum_{i=1}^{N}
    \left( y_i - \bar{y} \right)
    \left( \hat{y}_i - \bar{\hat{y}} \right)
    }{
    \sqrt{
    \sum_{i=1}^{N}
    \left( y_i - \bar{y} \right)^2
    }
    \sqrt{
    \sum_{i=1}^{N}
    \left( \hat{y}_i - \bar{\hat{y}} \right)^2
    }
    } ,
    \label{eq:correlation}
\end{equation}
where $\bar{y}$ and $\bar{\hat{y}}$ are the mean values of the observed and predicted peak soft X-ray fluxes, respectively. 
The CC measures the association between the observed and predicted peak fluxes, with a value closer to 1 indicating stronger agreement. 

Although the main objective of this study is peak flux prediction, we additionally evaluate the classification performance between C-class and $\geq$M-class flares as a secondary metric. 
We compute the True Skill Statistic (TSS; \citealp{ALLOUCHE_2006}) and F1 score \citep{vanRijsbergen1979} to quantify this supplementary classification performance. 
In this binary classification problem, $\geq$M-class flares are treated as positive events and C-class flares are treated as negative events; therefore, True Positives (TP) correspond to $\geq$M-class flares correctly predicted as $\geq$M-class, False Positives (FP) correspond to C-class flares incorrectly predicted as $\geq$M-class, True Negatives (TN) correspond to C-class flares correctly predicted as C-class, and False Negatives (FN) correspond to $\geq$M-class flares incorrectly predicted as C-class.

The TSS is defined as
\begin{equation}
    \mathrm{TSS}
    =
    \frac{\mathrm{TP}}{\mathrm{TP}+\mathrm{FN}}
    -
    \frac{\mathrm{FP}}{\mathrm{FP}+\mathrm{TN}} .
    \label{eq:tss}
\end{equation}
TSS ranges from $-1$ to $1$, where $1$ represents perfect classification, while $-1$ indicates performance worse than random prediction. It is useful for imbalanced event prediction problems.

The F1 score is defined as the harmonic mean of precision and recall:
\begin{equation}
\begin{gathered}
    \mathrm{F1}
    =
    2 \times
    \frac{
    \mathrm{Precision} \times \mathrm{Recall}
    }{
    \mathrm{Precision} + \mathrm{Recall}
    }, \\
    \mathrm{Precision}
    =
    \frac{\mathrm{TP}}{\mathrm{TP}+\mathrm{FP}}, \\
    \mathrm{Recall}
    =
    \frac{\mathrm{TP}}{\mathrm{TP}+\mathrm{FN}} .
\end{gathered}
\label{eq:f1}
\end{equation}







\subsection{Uncertainty Estimation} \label{subsec:uncertainty_evaluation}
We estimate the predictive uncertainty following the uncertainty quantification procedure of \citet{2025ApJS..280...50J}. This procedure combines Monte Carlo dropout (MC dropout), motivated by the Bayesian interpretation of dropout proposed by \citet{pmlr-v48-gal16}, with an adaptive noise estimation approach based on \citet{8215650}. For a given input data, the model is evaluated $T$ times with dropout enabled during inference. The variance $\sigma_{\rm model}^{2}$, which quantifies the model uncertainty, is approximated by the sampling variance of the MC dropout predictions:
\begin{equation}
\sigma_{\rm model}^{2}
=
\frac{1}{T}
\sum_{t=1}^{T}
\left(
\hat{y}^{(t)}-\bar{\hat{y}}
\right)^2,
\qquad
\bar{\hat{y}}
=
\frac{1}{T}
\sum_{t=1}^{T}
\hat{y}^{(t)} ,
\end{equation}
where $\hat{y}^{(t)}$ is the predicted peak soft X-ray flux from the $t$th stochastic forward pass, $\bar{\hat{y}}$ is the mean prediction over $T$ stochastic forward passes, and $T$ is the number of stochastic forward passes. 
In this study, $T$ is set to 100, and the dropout rate of 0.1 is applied to the final fully connected layer of the decoder.

In addition, we compute the noise variance, $\sigma_{\rm noise}^{2}$, which represents the inherent noise estimated from the validation data:
\begin{equation}
\sigma_{\rm noise}^{2}
=
\frac{1}{V}
\sum_{v=1}^{V}
\left(
y_v - f(x_v)
\right)^2 ,
\end{equation}
where $x_v$ is the $v$th input profile in the validation data, $y_v$ is the corresponding observed peak soft X-ray flux, $f(x_v)$ is the model prediction for $x_v$, and $V$ is the number of validation samples.
In our four-fold cross-validation experiment, a separate validation set is not used. Therefore, the noise variance is estimated separately for each test fold. 
When evaluating one test fold, the prediction results from the other test folds are used as validation data for computing $\sigma_{\rm noise}^{2}$. 
The final noise variance is then obtained by averaging the noise variances estimated for the four test folds.

The total variance, $\sigma_{\rm total}^{2}$, is obtained by combining the model variance and the noise variance: 
\begin{equation}
\sigma_{\rm total}^{2}
=
\sigma_{\rm model}^{2}
+
\sigma_{\rm noise}^{2}.
\end{equation}
The model, noise, and total uncertainties are finally expressed as $\sigma_{\rm model}$, $\sigma_{\rm noise}$, and $\sigma_{\rm total}$.

After the prediction uncertainties are determined, an approximate $\alpha$-level prediction interval can be constructed using a given uncertainty estimate:
\begin{equation}
\left[
\bar{\hat{y}} - z_{\alpha/2} \times \sigma_u,
\bar{\hat{y}} + z_{\alpha/2} \times \sigma_u
\right],
\end{equation}
where $\sigma_u$ denotes the uncertainty estimate used to define the interval. $\sigma_u$ can be $\sigma_{\rm model}$, $\sigma_{\rm noise}$, or $\sigma_{\rm total}$. Here, $z_{\alpha/2}$ is the upper $\alpha/2$ quantile of the standard normal distribution. We set $z_{\alpha/2}=2$, which approximately corresponds to a 95\% prediction interval \citep{2014bda..book.....G}.

To evaluate the reliability of the estimated uncertainty, we compute the Empirical Coverage (EC), which measures the fraction of observed values that fall within the corresponding prediction interval. For a given uncertainty estimate $\sigma_u$, the EC is defined as
\begin{equation}
{\rm EC}_u
=
\frac{1}{N}
\sum_{i=1}^{N}
I
\left(
y_i \in
\left[
\bar{\hat{y}}_i - z_{\alpha/2}\sigma_{u,i},
\bar{\hat{y}}_i + z_{\alpha/2}\sigma_{u,i}
\right]
\right)
\times 100\% ,
\end{equation}
where $N$ is the total number of test samples from the four-fold cross-validation results and $I(\cdot)$ is the indicator function, which is equal to 1 if the condition is satisfied and 0 otherwise. Here, $\sigma_u$ can be $\sigma_{\rm model}$, $\sigma_{\rm noise}$, or $\sigma_{\rm total}$. Unlike the other evaluation metrics, EC is computed from the full set of test samples from the four-fold cross-validation results, rather than from the average of fold-wise scores.

\section{Results} \label{sec:results}

\subsection{Flare Rise Time Analysis} \label{subsec:flare_risetime_analysis}

\begin{table}[!htb]
\centering
\caption{
Solar flare rise time statistics for C-, M-, and X-class flares from 1997 to 2024.
}
\label{tab:rise_time_statistics}
\hspace{-0.3in}\begin{tabular}{lrrr}
\hline
\textbf{Class} & 
\textbf{Number of events} & 
\textbf{Mean rise time} & 
\textbf{Std. dev. of rise time} \\
 &  & \textbf{(min.)} & \textbf{(min.)} \\
\hline
C & 23064 & 8.1  & 3.7 \\
M & 2880  & 12.4 & 6.0 \\
X & 197   & 17.5 & 8.2 \\
\hline
\end{tabular}\hspace{+0.3in}
\end{table}

Table~\ref{tab:rise_time_statistics} shows the mean and standard deviation of solar flare rise time calculated from the GOES solar flare event list from 1997 to 2024. The rise time is defined as the time interval from the flare onset to the flare peak. To reduce the influence of outliers, the upper and lower 10\% of events based on rise time were excluded separately for each flare class. 
These results indicate that the RMN strategy can be applied before the peak time even for many C-class flares, although C-class flares have the shortest rise times among the three classes. 
For stronger flares, the longer rise time allows peak flux prediction to be performed over a longer time window before the flare reaches its maximum.

\subsection{Model Performance for Regression} \label{subsec:peak_performance}

\begin{table}[!htb]
\centering
\caption{
Model performance and EC for the $\geq$C-, $\geq$M-, and X-class flare groups from three minutes after flare onset to the observed flare peak.
}
\label{tab:performance_by_class}
\hspace{-0.3in}\begin{tabular}{lccccc}
\hline
\textbf{Flare class} & \textbf{RMSE} & \textbf{PE} &
\textbf{$\mathrm{EC}_{total}$} &
\textbf{$\mathrm{EC}_{model}$} &
\textbf{$\mathrm{EC}_{noise}$} \\
\hline
$\geq$C & 0.26 & 3.11\%  & 96.1\% & 82.1\% & 95.2\% \\
$\geq$M & 0.45 & 5.59\%  & 91.7\% & 63.4\% & 90.5\% \\
X       & 0.87 & 12.76\% & 83.8\% & 40.7\% & 82.9\% \\
\hline
\end{tabular}\hspace{+0.3in}
\end{table}

Table~\ref{tab:performance_by_class} presents the regression performance of the model for different flare class groups. The performance metrics are computed using all predictions made at one-minute intervals from three minutes after flare onset to the observed flare peak.
The model achieves the highest accuracy for the $\geq$C-class group, with an RMSE of 0.26 and PE of 3.30\%.
The errors increase as flares become more intense, with the RMSE increasing to 0.45 for the $\geq$M-class group and 0.87 for the X-class group.
This result suggests that predicting the peak flux of more intense flares is more challenging, probably because such events are much less frequent and have more dynamically evolving X-ray profiles.
The EC based on total uncertainty also decreases from 96.1\% for the $\geq$C-class group to 91.7\% for the $\geq$M-class group and 83.8\% for the X-class group. 
This decrease may be related to the limited number of strong-flare events and the greater difficulty of predicting their peak fluxes. 
The EC values based on model uncertainty alone are lower than those based on total uncertainty, particularly for the X-class group, with values of 82.1\%, 63.4\%, and 40.7\% for the $\geq$C-, $\geq$M-, and X-class groups, respectively. 
In contrast, the EC values based on noise uncertainty are close to those based on total uncertainty, with values of 95.2\%, 90.5\%, and 82.9\%, indicating that the data-derived noise component makes a major contribution to the final prediction uncertainty.

\begin{figure}[!htb]
    \centering
    \begin{subfigure}{0.32\columnwidth}
        \centering
        \includegraphics[width=\linewidth]{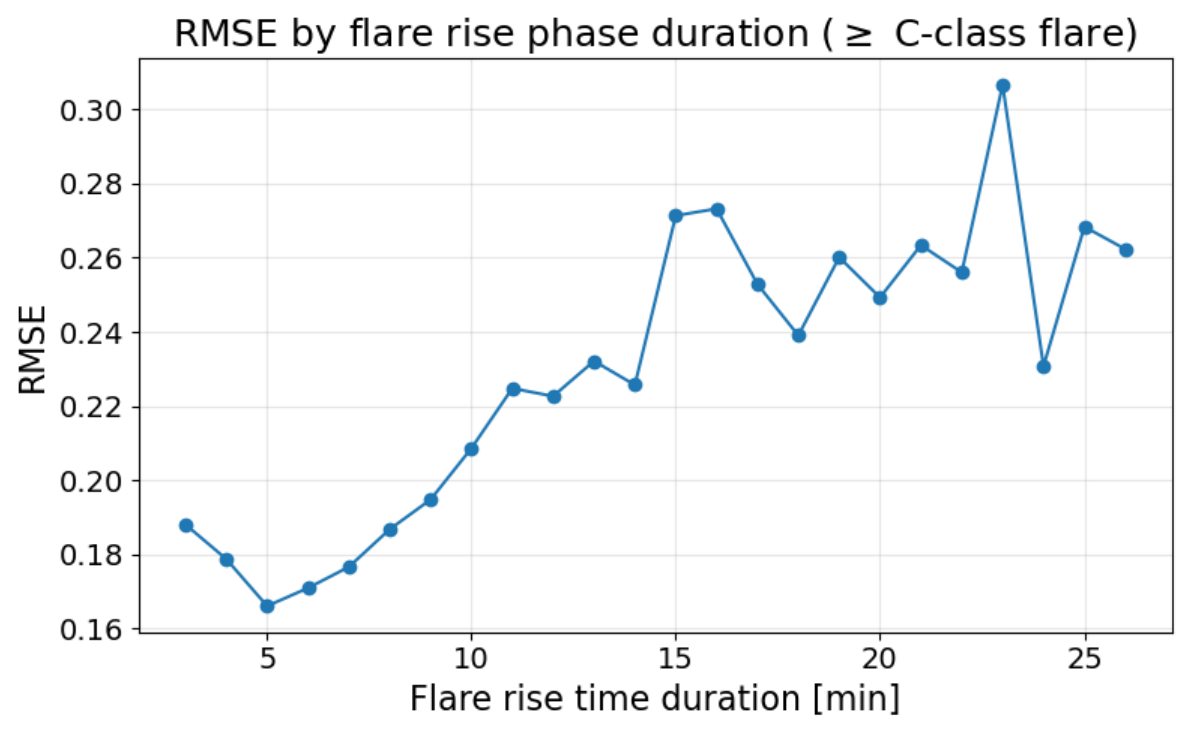}
        \caption{$\geq$ C-class flare}
        \label{fig:fig1}
    \end{subfigure}
    \hfill
    \begin{subfigure}{0.32\columnwidth}
        \centering
        \includegraphics[width=\linewidth]{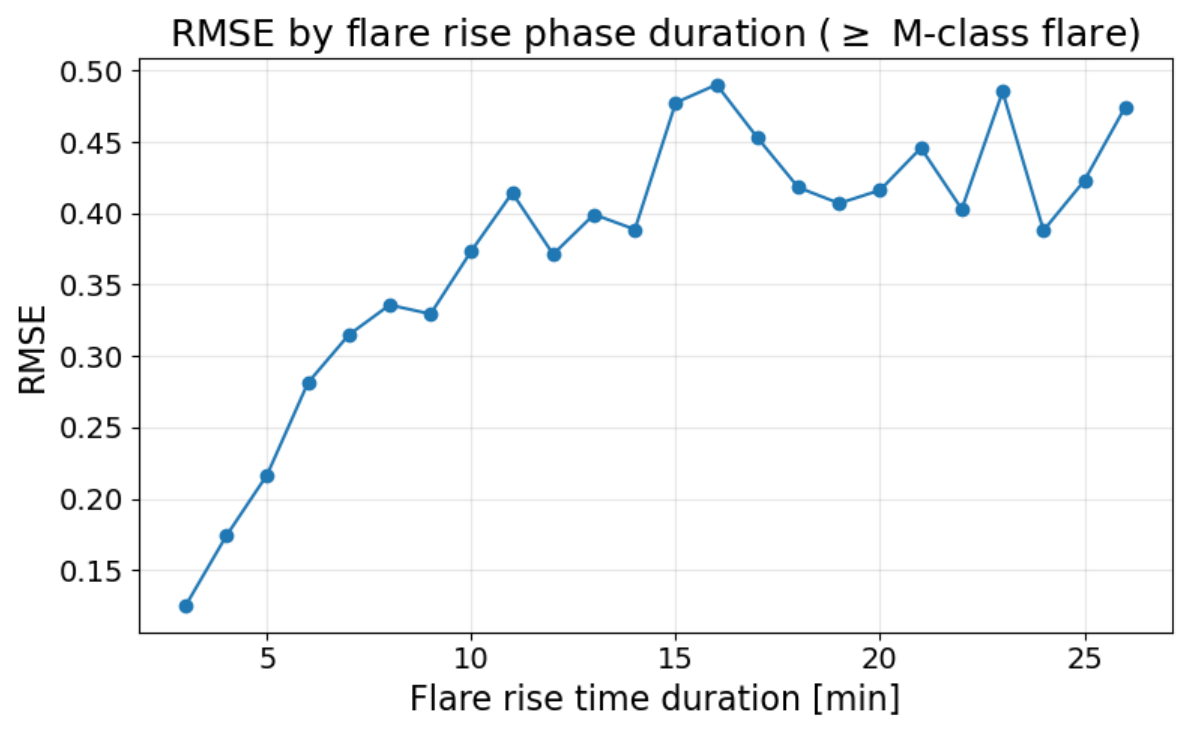}
        \caption{$\geq$ M-class flare}
        \label{fig:fig2}
    \end{subfigure}
    \hfill
    \begin{subfigure}{0.32\columnwidth}
        \centering
        \includegraphics[width=\linewidth]{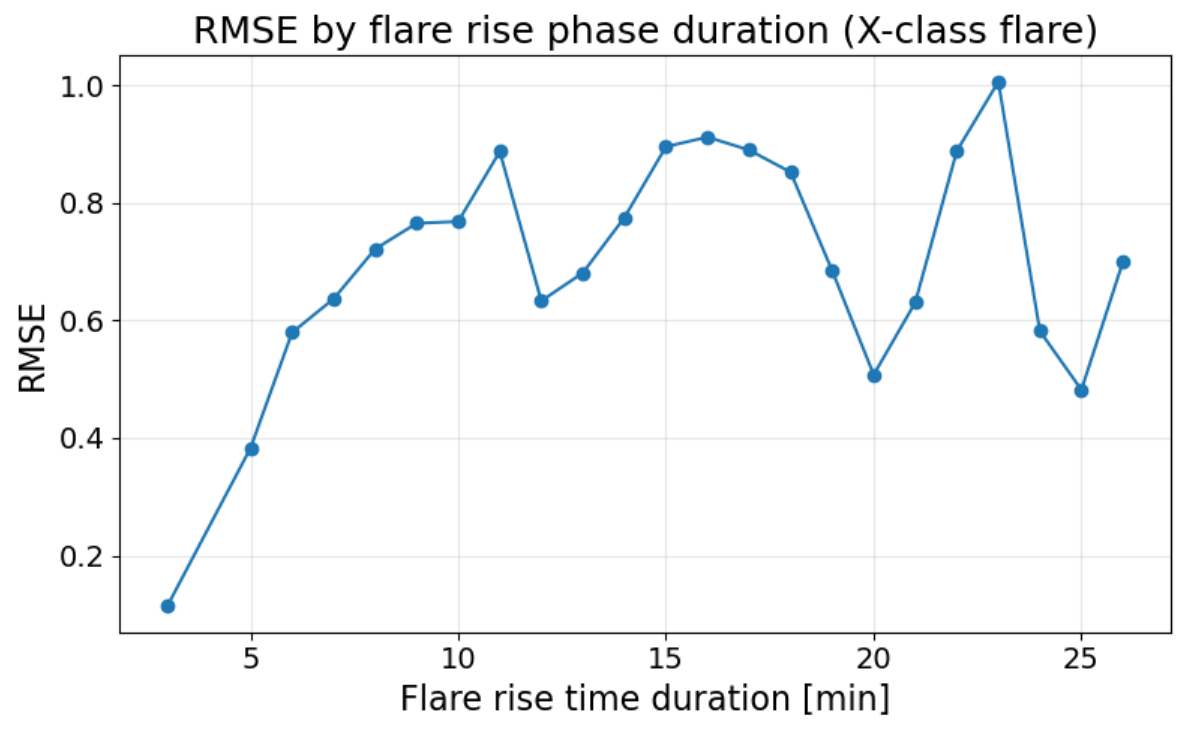}
        \caption{X-class flare}
        \label{fig:fig3}
    \end{subfigure}

    \caption{Model's RMSE by flare rise time duration for different flare classes.}
    \label{fig:three_figures}
\end{figure}

To further examine the factors associated with variations in prediction accuracy, we analyze the model performance as a function of flare rise time duration.
Figure~\ref{fig:three_figures} shows the model RMSE for flare events grouped according to their flare rise time duration in the $\geq$C-, $\geq$M-, and X-class flare groups. 
To reduce the influence of outliers, flares in the upper 10\% of rise time duration were excluded separately for each flare class group. 
In all three groups, the RMSE generally increases with rise time duration, indicating that the model performs better for flares with shorter rise time.
Flares with shorter rise times reach a substantial fraction of their peak flux more rapidly, allowing the observed X-ray profile to provide a clearer indication of the eventual peak. 
For longer-rise events, the flux remains farther below the final peak for a longer period, making the peak-flux prediction less accurate.
In addition to this overall dependence on rise time duration, the X-class results show relatively large fluctuations between rise time groups. 
This is likely due to the small number of X-class events in each group, which makes the RMSE more sensitive to a few events with large prediction errors.

\begin{figure}[!htb]
\centering
\includegraphics[width=1\linewidth]{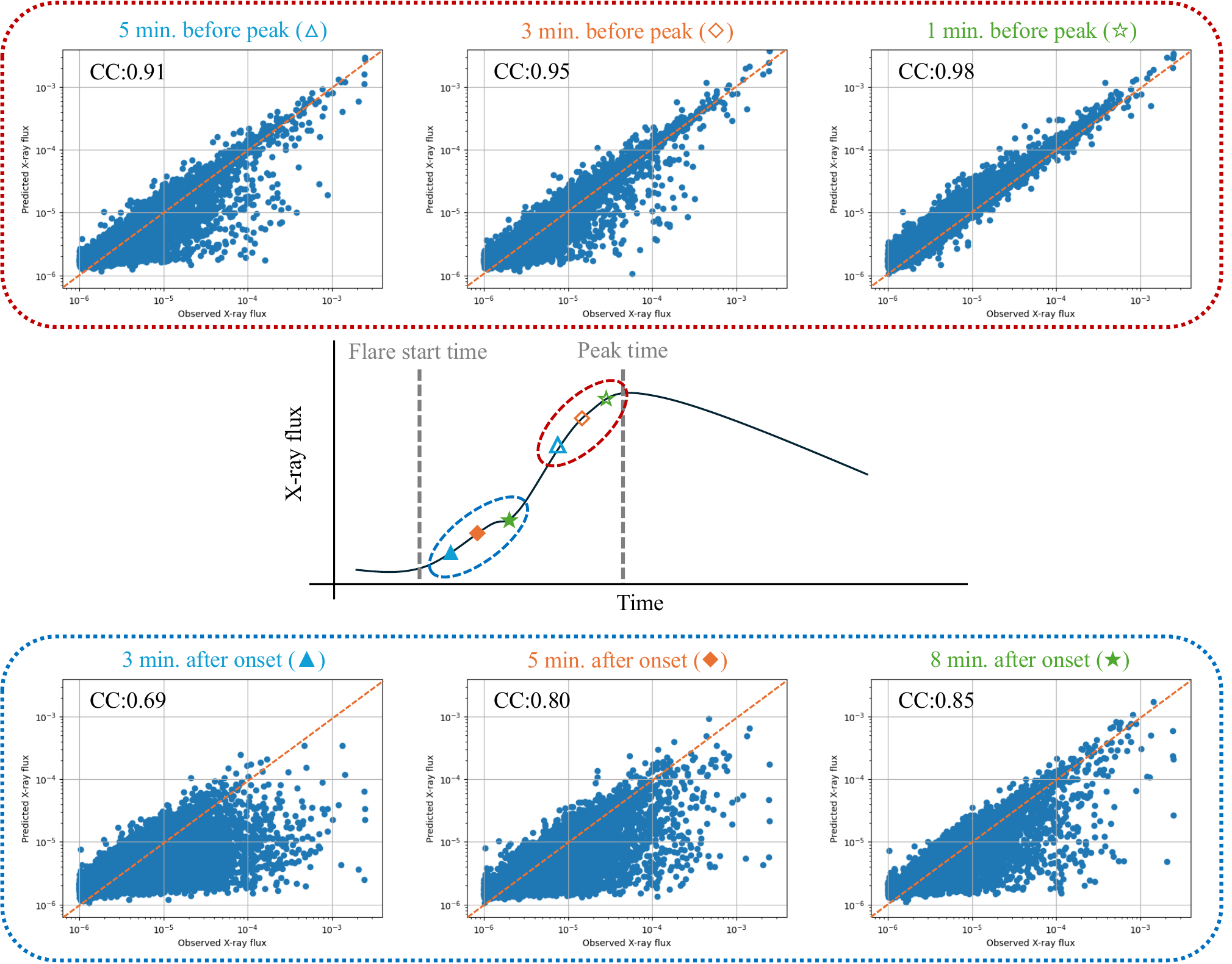}
\caption{
Scatter plots of observed and predicted peak soft X-ray fluxes for the $\geq$C-class flare group at different prediction times.
(Top) Prediction results at the late rise phase before the flare peak time.
(Middle) An example of X-ray flux profile during a flare event, with symbols marking the prediction times corresponding to the scatter plots. 
(Bottom) Prediction results at the early rise phase after the flare onset time.
The dashed line in each scatter plot represents the one-to-one line.
}
\label{fig:scatter}
\end{figure}

\begin{table*}[!htb]
\centering
\caption{
Model performance as a function of prediction time relative to flare onset and flare peak.
Positive (+) times indicate minutes after flare onset, while negative (-) times indicate minutes before flare peak.
}
\label{tab:performance_by_predictiontime}
\resizebox{\textwidth}{!}{%
\hspace{-0.8in}\begin{tabular}{lcccccc c cccccc}
\hline
\multicolumn{14}{c}{\textbf{$\geq$ C-class flare}} \\
\hline
& \multicolumn{6}{c}{\textbf{After flare onset}}
& 
& \multicolumn{6}{c}{\textbf{Before flare peak}} \\
\textbf{Metric}
& \textbf{+3 min} & \textbf{+4 min} & \textbf{+5 min} & \textbf{+6 min}
& \textbf{+7 min} & \textbf{+8 min} & $\cdots$
& \textbf{-5 min} & \textbf{-4 min} & \textbf{-3 min}
& \textbf{-2 min} & \textbf{-1 min} & \textbf{Peak} \\
\hline
RMSE
& 0.32 & 0.28 & 0.27 & 0.27 & 0.26 & 0.25 & $\cdots$
& 0.20 & 0.18 & 0.17 & 0.16 & 0.16 & 0.12 \\
PE
& 4.24\% & 3.67\% & 3.47\% & 3.38\% & 3.26\% & 3.21\% & $\cdots$
& 2.69\% & 2.53\% & 2.45\% & 2.42\% & 2.47\% & 1.81\% \\
\hline

\multicolumn{14}{c}{\textbf{$\geq$ M-class flare}} \\
\hline
& \multicolumn{6}{c}{\textbf{After flare onset}}
& 
& \multicolumn{6}{c}{\textbf{Before flare peak}} \\
\textbf{Metric}
& \textbf{+3 min} & \textbf{+4 min} & \textbf{+5 min} & \textbf{+6 min}
& \textbf{+7 min} & \textbf{+8 min} & $\cdots$
& \textbf{-5 min} & \textbf{-4 min} & \textbf{-3 min}
& \textbf{-2 min} & \textbf{-1 min} & \textbf{Peak} \\
\hline
RMSE
& 0.67 & 0.59 & 0.54 & 0.51 & 0.48 & 0.45 & $\cdots$
& 0.33 & 0.28 & 0.23 & 0.17 & 0.13 & 0.09 \\
PE
& 9.95\% & 8.42\% & 7.43\% & 6.77\% & 6.28\% & 5.72\% & $\cdots$
& 3.97\% & 3.41\% & 2.98\% & 2.49\% & 2.15\% & 1.45\% \\
\hline

\multicolumn{14}{c}{\textbf{X-class flare}} \\
\hline
& \multicolumn{6}{c}{\textbf{After flare onset}}
& 
& \multicolumn{6}{c}{\textbf{Before flare peak}} \\
\textbf{Metric}
& \textbf{+3 min} & \textbf{+4 min} & \textbf{+5 min} & \textbf{+6 min}
& \textbf{+7 min} & \textbf{+8 min} & $\cdots$
& \textbf{-5 min} & \textbf{-4 min} & \textbf{-3 min}
& \textbf{-2 min} & \textbf{-1 min} & \textbf{Peak} \\
\hline
RMSE
& 1.36 & 1.20 & 1.09 & 1.02 & 0.96 & 0.89 & $\cdots$
& 0.55 & 0.46 & 0.40 & 0.20 & 0.13 & 0.11 \\
PE
& 24.93\% & 21.48\% & 18.75\% & 16.58\% & 15.24\% & 13.71\% & $\cdots$
& 7.11\% & 5.91\% & 5.24\% & 3.29\% & 2.80\% & 2.26\% \\
\hline
\end{tabular}\hspace{+0.4in}
}%
\end{table*}

Table~\ref{tab:performance_by_predictiontime} presents the prediction performance evaluated separately at each prediction time for the $\geq$C-, $\geq$M-, and X-class flare groups. 
The prediction accuracy improves as the prediction time moves from the early phase after flare onset toward the flare peak. For all flare groups, the RMSE and PE decrease with prediction time. 
For the $\geq$C-class group, the RMSE decreases from 0.32 at 3 min after flare onset to 0.12 at the peak time. 
Similar improvements are also found for the $\geq$M- and X-class groups.
For the $\geq$M-class group, the RMSE decreases from 0.67 to 0.09, and the PE decreases from 9.95\% to 1.45\%. 
For the X-class group, the RMSE decreases from 1.36 to 0.16, and the PE decreases from 24.93\% to 2.26\%. 
These results indicate that early-stage prediction is more challenging for stronger flares, but the RMN strategy can improve the nowcasting performance as the flare approaches its peak flux. 
The high performance near the flare peak time suggests that this approach may also provide useful information for assessing whether an ongoing flare is still increasing in intensity or is close to reaching its peak flux.

Figure~\ref{fig:scatter} shows scatter plots of the observed and predicted peak soft X-ray fluxes for the $\geq$C-class flare group at different prediction times.
The top panels show late phase prediction results obtained 5, 3, and 1 min before the flare peak, while the bottom panels show early phase prediction results obtained 3, 5, and 8 min after flare onset.
The middle panel shows an example X-ray flux profile during a flare event, with symbols indicating the prediction times corresponding to the scatter plots.
The results show underprediction during the early rise phase with a CC of 0.69 at 3 min after flare onset. At this stage, the observed X-ray flux profile contains limited information for distinguishing the eventual peak intensity of strong flares from that of the more common, weaker events. 
As the prediction time approaches the flare peak, the scatter points become more tightly concentrated around the one-to-one line, and the CC increases to 0.98 at 1 min before the flare peak. 
The mean prediction error ($\frac{1}{N}\sum_{i=1}^{N}\left(\log_{10}F_{\mathrm{pred},i}-\log_{10}F_{\mathrm{obs},i}\right)$) shows a stronger negative bias for more intense flares during the early rise phase. At 3 min after flare onset, the mean prediction errors are $-0.02$, $-0.47$, and $-1.25$ for the $\geq$C-, $\geq$M-, and X-class flares, respectively. The negative bias for stronger flares is reduced as the prediction time approaches the flare peak, with mean prediction errors of $0.13$, $0.08$, and $0.06$, respectively, at 1 min before the peak.
These scatter plots are consistent with the time-dependent performance shown in Table~\ref{tab:performance_by_predictiontime} and support the effectiveness of the RMN strategy for peak flux nowcasting during the flare rise phase.

\subsection{Derived Flare-Class Performance} \label{subsec:class_performance}


The predicted peak soft X-ray flux achieved a TSS of 0.68 and an F1 score of 0.77 when converted into flare classes using the NOAA classification thresholds.
The positive TSS value shows that the RMN strategy has meaningful discrimination capability by considering both TP and FP rates, while the F1 score indicates a reasonable balance between precision and recall for $\geq$M-class flare detection.
These results indicate that the predicted peak soft X-ray flux can distinguish $\geq$M-class flares from C-class flares with reasonable skill.

\begin{table*}[!htb]
\centering
\caption{
Classification performance for distinguishing C-class and $\geq$M-class flares as a function of prediction time relative to flare onset and flare peak. The $\geq$M-class flares are treated as the positive class. Positive (+) times indicate minutes after flare onset, while negative (-) times indicate minutes before flare peak.
}
\label{tab:classification_performance_by_predictiontime}
\hspace{-0.3in}\resizebox{\textwidth}{!}{%
\begin{tabular}{@{}l*{6}{c}c*{6}{c}@{}}
\hline
& \multicolumn{6}{c}{\textbf{After flare onset}}
& 
& \multicolumn{6}{c}{\textbf{Before flare peak}} \\
\textbf{Metric}
& \textbf{+3 min} & \textbf{+4 min} & \textbf{+5 min}
& \textbf{+6 min} & \textbf{+7 min} & \textbf{+8 min}
& $\cdots$
& \textbf{-5 min} & \textbf{-4 min} & \textbf{-3 min}
& \textbf{-2 min} & \textbf{-1 min} & \textbf{Peak} \\
\hline
TSS
& 0.36 & 0.49 & 0.56 & 0.60 & 0.64 & 0.67
& $\cdots$
& 0.76 & 0.81 & 0.84 & 0.89 & 0.93 & 0.94 \\
F1
& 0.50 & 0.61 & 0.68 & 0.71 & 0.74 & 0.76
& $\cdots$
& 0.81 & 0.83 & 0.83 & 0.84 & 0.84 & 0.89 \\
\hline
\end{tabular}%
}\hspace{+0.3in}
\end{table*}

Table~\ref{tab:classification_performance_by_predictiontime} presents the classification performance for C-class and $\geq$M-class flares at different prediction times. 
Both the TSS and F1 score increase as the prediction time approaches the flare peak. 
The TSS increases from 0.36 at 3 min after flare onset to 0.94 at the flare peak, while the F1 score increases from 0.50 to 0.89 over the same period. 
These results indicate that the classification performance is lower during the early rise phase, but improves as the flare approaches its peak flux.

Considering the trade-off between prediction accuracy and lead time, 5 min after flare onset can be considered a reasonable operational prediction time for the present model. 
At this time, the model uses additional information from the early rise phase, and both the peak-flux regression and $\geq$M-class flare classification performances improve relative to those at the earliest prediction time. This timing also retains lead time before the flare peak for many events, providing a practical condition for applying the RMN strategy to ongoing flares.

\subsection{Near-real-time flare nowcast}

We have implemented the models developed in this study in a publicly accessible near-real-time flare nowcasting system\footnote{\url{https://kwyi.github.io/Flare_Nowcasting/}}. The system obtains near-real-time one-minute GOES XRS observations provided by NOAA\footnote{\url{https://services.swpc.noaa.gov/json/goes/primary/xrays-6-hour.json}} and applies the flare-detection algorithm described in the GOES-R XRS L2 Data Users Guide to determine the status of solar X-ray flare events. 
Once an ongoing flare is detected, each of the four cross-validation models estimates its eventual peak soft X-ray flux using the preceding 60 min of observations. 
The mean of the four model outputs is used as the final nowcasting result.

\begin{figure}
\centering
\includegraphics[width=1.0\linewidth]{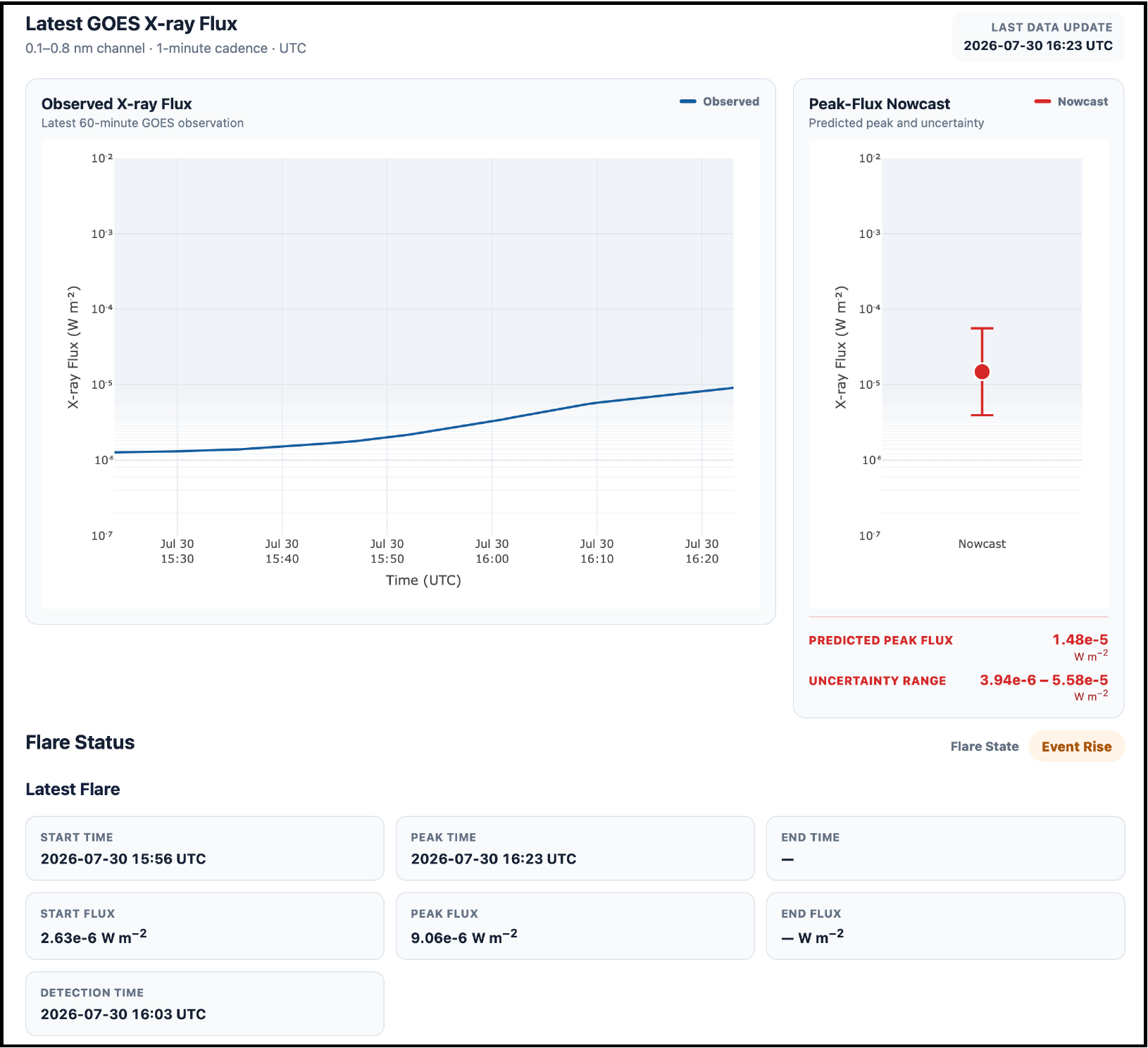}
\caption{
Example of the near-real-time flare nowcasting website. The blue line shows the latest 60 minutes of one-minute-averaged GOES 0.1--0.8~nm X-ray flux. The red circle indicates the predicted peak flux of the ongoing flare, with the prediction interval.
}
\label{fig:nowcasting_sample}
\end{figure}

Figure~\ref{fig:nowcasting_sample} shows an example displayed on the near-real-time flare nowcasting website. 
The blue curve represents the one-minute-averaged solar X-ray flux in the 0.1--0.8~nm channel during the preceding hour. 
When an ongoing flare is detected, the model continuously estimates its eventual peak flux, which is shown by the red circle with the prediction interval based on the total uncertainty with $z_{\alpha/2}=2$.
The website also provides the start, peak, and end information for the most recently detected flare, together with the flare detection time determined by the flare-detection algorithm.
While a flare is ongoing, the peak time and peak flux are continuously updated using the latest available observations.

\begin{figure}
\centering
\includegraphics[width=1.0\linewidth]{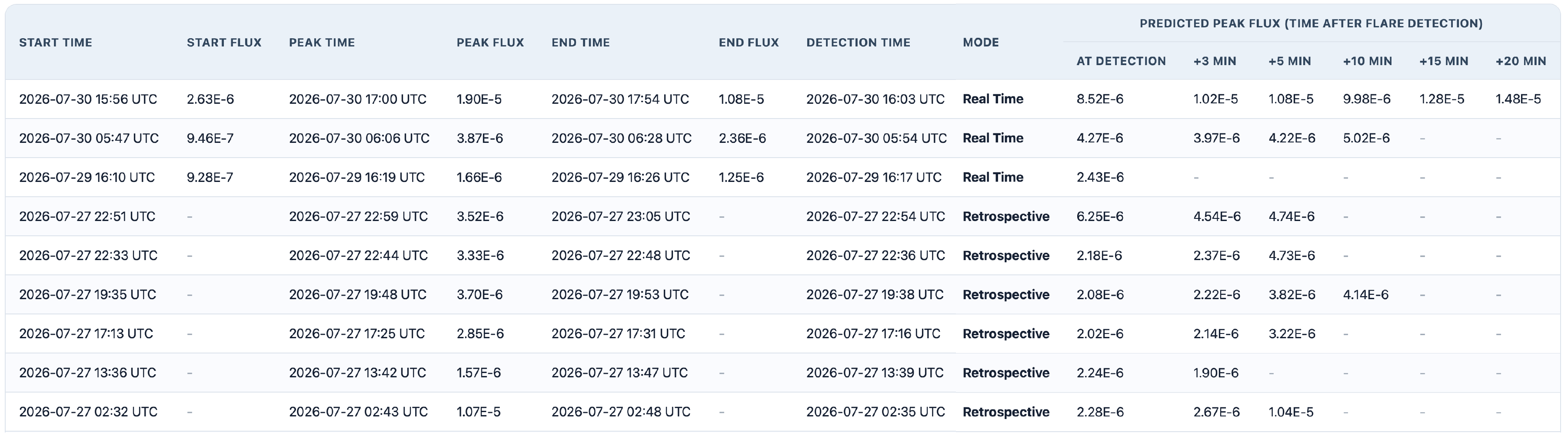}
\caption{
Nowcasting archive for retrospective flares in 2025-2026. 
The flare start, peak, and end times and fluxes are listed. 
The detection time represents when the flare is identified by the flare-detection algorithm.
Real time mode indicates predictions generated during real-time operation, whereas retrospective mode indicates predictions generated for past flare events. Predictions at 0, 3, 5, 10, 15, and 20 min after detection are recorded.
}
\label{fig:nowcasting_list}
\end{figure}

To retain a traceable record of the model outputs and enable subsequent performance evaluation, the nowcasting results are listed and displayed, as shown in Figure~\ref{fig:nowcasting_list}. 
In addition to results generated during near-real-time operation, the list contains retrospective nowcasting results for flares that occurred between 2025 January 1 and 2026 July 27.
Each retrospective flare was assumed to have been detected three minutes after its cataloged onset time.
The model generated peak flux predictions from this assumed detection time to the observed peak time.
The live display and the nowcasting archive provide a public demonstration of the proposed nowcasting system and an event list for evaluating its performance under operationally realistic conditions.



\begin{table}[!htb]
\centering
\caption{
Retrospective model performance and EC for the $\geq$C-, $\geq$M-, and X-class flare groups.
}
\label{tab:performance_by_class_2025-2026}
\hspace{-0.3in}\begin{tabular}{lccccc}
\hline
\textbf{Flare class} & \textbf{RMSE} & \textbf{PE} &
\textbf{$\mathrm{EC}_{total}$} &
\textbf{$\mathrm{EC}_{model}$} &
\textbf{$\mathrm{EC}_{noise}$} \\
\hline
$\geq$C & 0.25 & 3.27\%  & 96.2\% & 84.7\% & 95.3\% \\
$\geq$M & 0.45 & 5.87\%  & 91.7\% & 62.2\% & 91.0\% \\
X       & 0.84 & 12.53\% & 84.2\% & 44.7\% & 82.3\% \\
\hline
\end{tabular}\hspace{+0.3in}
\end{table}

The retrospective nowcasting archive contains 3,317 flare events, comprising 2,914 C-class flares (87.9\%), 379 M-class flares (11.4\%), and 24 X-class flares (0.7\%).
Compared with the 1997--2024 training and validation dataset, the retrospective dataset shows a stronger concentration of C-class flares relative to M- and X-class events.
Table~\ref{tab:performance_by_class_2025-2026} summarizes the overall prediction performance calculated using all nowcasts from the assumed flare detection time to the observed peak time.
The overall performance of the retrospective evaluation is broadly comparable to that obtained from the 1997--2024 evaluation analysis.

\section{Conclusion and Discussion} \label{sec:conclusion_discussion}
In this study, we present the RMN strategy for nowcasting the peak soft X-ray flux of ongoing solar flares using real-time GOES X-ray observations, machine learning model, and NOAA flare detection criteria. 
The main objective of this study is to quantitatively evaluate the capability of the RMN strategy for peak flux nowcasting of ongoing flares under operationally realistic conditions, where prediction starts from the earliest time at which a flare can be identified in real time.
For this, we used the GOES flare event list and the corresponding one-minute GOES 0.1--0.8 nm soft X-ray flux data for C-, M-, and X-class flares observed from 1997 to 2024.
We evaluated the nowcasting performance using RMSE, PE, and CC for peak flux prediction, TSS and F1 score for C-class versus $\geq$M-class classification, and EC with MC dropout for uncertainty estimation.

The major results of this study are as follows. 

%

\begin{itemize}
\item First, the proposed model successfully predicts the peak soft X-ray flux of ongoing flares. 
The RMSE and PE are 0.26 and 3.11\% for the $\geq$C-class flare group; 0.45 and 5.59\% for the $\geq$M-class flare group; and 0.87 and 12.76\% for the X-class flare group, respectively. 
The predicted peak flux also classifies C-class and $\geq$M-class flares with a TSS of 0.68 and an F1 score of 0.77.
\item Second, the model performance depends on flare rise time and prediction time. The $\geq$C-, $\geq$M-, and X-class flare groups show larger RMSE values for longer rise events than for shorter rise events. The RMSE and PE decrease as the prediction time approaches the flare peak.
\item Third, the uncertainty analysis shows that the EC based on total uncertainty remains high across all flare groups, and decreases for stronger flares. The lower EC for stronger flares is consistent with the greater difficulty of predicting their peak fluxes. The comparison between three uncertainty values shows that the noise uncertainty makes a larger contribution to the total prediction uncertainty than the model uncertainty.
\end{itemize}

The proposed models have also been implemented in a publicly accessible near-real-time flare nowcasting website. 
The website applies the proposed models to GOES XRS observations and displays the peak flux nowcasting result and its uncertainty for an ongoing flare.
The website also provides an archive of model nowcastging results, allowing the prediction performance to be evaluated across individual flare events under operationally realistic conditions.

The present study evaluates peak flux nowcasting under operationally realistic conditions by retaining the observed flare-class distribution.
Weak flares occur much more frequently than strong flares, resulting in an imbalanced dataset problem with a limited number of strong flares in both the training and test datasets. 
During model development, we tested stratified undersampling, stratified oversampling, and weighted loss functions, but these approaches did not improve the model performance compared with training using the original flare-class distribution. 
The limited number of strong flares in the dataset can increase the statistical uncertainty of the performance metrics for stronger flare groups.


The relatively large errors for longer rise time events may be partly related to multiple-peak flare events. 
A long rise time can indicate that multiple flares overlap and are observed as a single event. 
In multiple peak flare events, predicting the peak flux during the early flare rise phase is especially difficult because the initial X-ray increase may not contain enough information about later enhancements.
In this study, multiple-peak flare events are included in the model evaluation to reflect realistic flare nowcasting conditions. 
These events may therefore contribute to the poorer performance for longer rise time events.

Early-stage prediction errors and the underestimation of strong flares may be reduced by incorporating additional real-time or near-real-time observations, which can provide additional information on flare evolution.
Solar microwave observations can serve as precursors of flare X-ray emission by providing information on nonthermal electron acceleration during the early phase of flares, and such data can be obtained in real time from instruments such as the Expanded Owens Valley Solar Array (EOVSA).
Magnetogram data can also provide information on the magnetic features of active regions, which may be useful for estimating the potential magnitude of a flare.
Although L2 magnetogram data are not generally available in real time, quick-look magnetograms can provide near-real-time information with a delay of about two hours. 
In future work, we will extend the RMN strategy by using additional data sources such as EOVSA microwave data and magnetogram data to improve flare peak flux nowcasting.


\begin{acknowledgments}
We thank the numerous team members who have contributed to the success of the GOES mission. We thank contributors to PyTorch,
Numpy and the Matplotlib open-source package. This research was supported by NASA grants 80NSSC24K0548, 80NSSC25K7708, and 80NSSC24M0174.  The authors gratefully acknowledge the computational resources and support provided by the Wulver high-performance computing cluster at the New Jersey Institute of Technology (NJIT).
\end{acknowledgments}

%



\bibliography{Reference}{}
\bibliographystyle{aasjournalv7}



\end{document}